\documentclass[reprint,twocolumn,notitlepage,prb,superscriptaddress,showpacs]{revtex4-2}
\usepackage{hyperref}
\usepackage[usenames,dvipsnames]{color}
\usepackage{amsmath}
\usepackage[english]{babel}
\usepackage{amssymb}
\usepackage{graphicx}
\usepackage{epstopdf}
\usepackage{subfig}
\usepackage{comment}
\usepackage{mathrsfs}
\usepackage{amsfonts}
\usepackage{tabularx}
\usepackage{multirow}

\usepackage[normalem]{ulem}

\begin{document}

\title{Influence of incommensurate structure on the elastic constants of crystalline Bi$_2$Sr$_2$CaCu$_2$O$_{8+\delta}$ by Brillouin light scattering spectroscopy}
\author{B. D. E. McNiven}
\affiliation{Department of Physics and Physical Oceanography, Memorial University of Newfoundland, St. John's, Newfoundland \& Labrador, Canada A1B 3X7} 
\author{J. P. F. LeBlanc}
\affiliation{Department of Physics and Physical Oceanography, Memorial University of Newfoundland, St. John's, Newfoundland \& Labrador, Canada A1B 3X7}
\author{G. T. Andrews}
\email{tandrews@mun.ca}
\affiliation{Department of Physics and Physical Oceanography, Memorial University of Newfoundland, St. John's, Newfoundland \& Labrador, Canada A1B 3X7} 

\date{\today}
\begin{abstract}
Brillouin light scattering spectroscopy was used to probe the room-temperature elasticity of crystalline high-$T_c$ superconductor Bi$_2$Sr$_2$CaCu$_2$O$_{8+\delta}$.  A complete set of best-estimate elastic constants was obtained using established relationships between acoustic phonon velocities and elastic constants along with a simple expression relating crystal elastic constant $C_{22}$ to the corresponding constants of the constituent incommensurate sublattices.  This latter relationship, which was derived and validated in the present work, has important implications for those studying incommensurate systems as it appears that it may be applied in its general form to any composite incommensurate crystal.  The results obtained in this work are also consistent with sublattice assignments of Bi$_2$Sr$_2$O$_4$ and CaCu$_2$O$_4$ reported in a previous Brillouin scattering study.
\end{abstract}

\maketitle

\section{Introduction}
Bi$_2$Sr$_2$CaCu$_2$O$_{8+\delta}$ (Bi-2212) is one of the most intensely studied high-$T_c$ superconductors yet its elastic properties have not been definitively characterized nor its acoustic phonon dynamics deeply probed.  Early studies of Bi-2212 elasticity focused on the use of Brillouin light scattering spectroscopy and ultrasonics techniques to determine acoustic phonon velocities and elastic moduli \cite{Wu1,Wu2,Wang1,Saint-Paul,Aleksandrov,Baumgart,Boekholt}.  Some elastic constants, however, remain unmeasured even at room temperature and the values of others uncertain due to factors such as the nature of approximations employed, sub-optimal crystal quality, and, in the case of Brillouin studies, extremely weak signals and the apparent lack of a reliable refractive index value from which to determine bulk acoustic phonon velocities from spectral peak frequency shifts \cite{Boekholt,Baumgart}.  Moreover, recent studies have uncovered interesting new acoustic phonon physics associated with the incommensurate structure of Bi-2212 \cite{McNiven2022,Etrillard2001,He2018,Merritt2019}, the investigation and understanding of which is in its infancy.  In particular, Brillouin \cite{McNiven2022} and neutron scattering \cite{Etrillard2001} studies revealed acoustic modes in excess of those expected for a typical commensurate crystal, calling into question mode assignments and consequently, elastic constants, reported in previous studies in which the incommensurability of Bi-2212 was not considered.  Furthermore, while both of these studies find that Bi-2212 is better classed as a composite rather than a modulated incommensurate crystal, they differ in constituent sublattice assignments, the former suggesting Bi$_2$Sr$_2$O$_4$ and CaCu$_2$O$_4$ \cite{McNiven2022} and the latter, Bi$_2$SrO$_{2+\delta}$ and SrCaCu$_2$O$_6$ \cite{Etrillard2001}.

This paper describes a series of Brillouin light scattering experiments aimed at addressing the paucity of experimental data on the elasticity of crystalline Bi-2212.  It does so by providing a set of best-estimate room-temperature elastic stiffness constants for crystalline Bi-2212 and its two constituent sublattices through the use of Brillouin peak frequency shift data and both established and newly-derived relationships between acoustic phonon velocities and elastic moduli. 

By reporting first values for previously unmeasured elastic constants, correcting or confirming early published values of others, and establishing connections between the incommensurate structure and elastic properties, this work provides new insights into the elasticity and acoustic phonon dynamics of Bi-2212.  The results deepen understanding of bonding in this material and in layered materials in general, but perhaps more importantly, aid in deciphering the role of electron-phonon coupling in high-$T_c$ superconductivity and in the development and refinement of models of elasticity and acoustic phonon dynamics in incommensurate systems. 

\section{Experimental Details}
Brillouin scattering experiments were performed under ambient conditions using a backscattering geometry on (001)-oriented flakes of Bi-2212 obtained from three larger parent single crystals with $T_c$ values of 78 K, 90 K, and 91 K. A single mode Nd:YVO$_4$ laser emitting at $\lambda = 532$ nm served as the incident light source. The laser beam was horizontally polarized using a half-wave plate and then passed through several attenuating filters to reduce the incident power to $\sim$10 mW in order to avoid sample damage. It was then focused onto the sample using a $f=5$ cm lens with $f/\# = 2.8$.  This lens was also used to collect and collimate backscattered light, which was subsequently focused onto the entrance pinhole of a six-pass tandem Fabry-Perot interferometer by a $f=40$ cm lens.    Further details on the set-up used in these experiments are provided in Ref. \cite{andr2018}.

Brillouin spectra were collected for light incident at angles ranging from $10^\circ \leq \theta_i \leq 75^\circ$, corresponding to $\sim5^\circ-28^\circ$ from the crystallographic c-axis.  Bulk acoustic phonon velocities, $V_B$, were determined directly from spectral peak frequency shifts using the well-known Brillouin equation $f_B = 2nV_B/\lambda$, where $f_B$ is the bulk phonon frequency shift and $n=2.0$ is the refractive index of Bi-2212 \cite{bozo1990,hwan2007,wang2012}.  Surface acoustic mode velocities were extracted from linear fits of the Brillouin equation for surface modes, $f_R = 2V_R \sin\theta_i / \lambda$, to experimental shift ({\it i.e.}, $f_R$) versus $\sin \theta_i$ data. In problematic cases in which two spectral peaks were very closely-spaced ({\it e.g.}, $QT_3$ and $QL_1$), Lorentzian functions were fitted to the peaks to facilitate extraction of accurate frequency shift values. 

\section{Results}
\subsection{Spectra}
\begin{figure}
    \centering
    \includegraphics[width=0.5\textwidth]{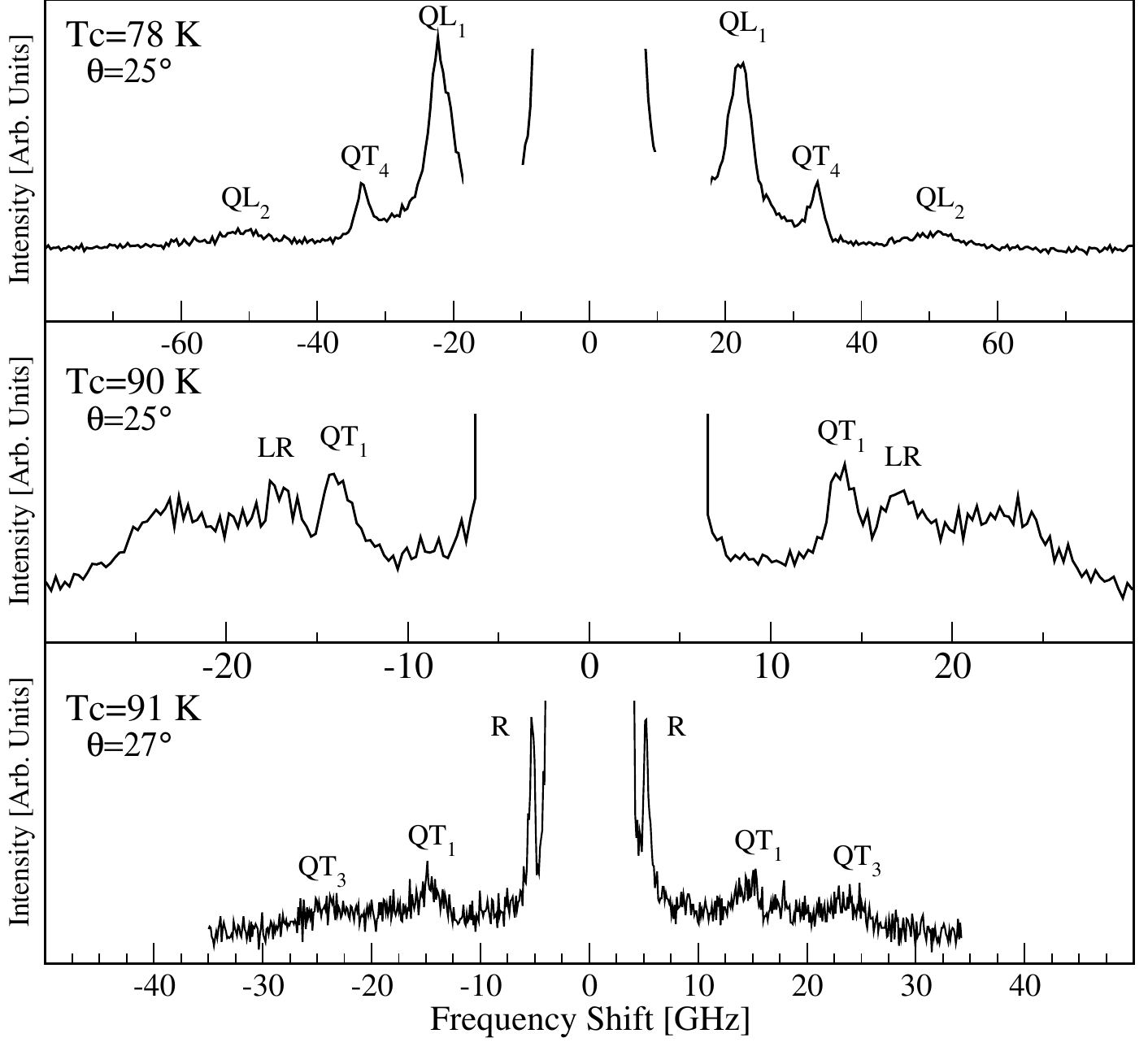}\\
    \caption{Room temperature Brillouin spectra of Bi-2212 single crystals with $T_c=78$ K, $90$ K, and $91$ K collected at indicated angles from the crystallographic $c$-axis.  The region close to the central elastic peak in the topmost spectrum was removed to eliminate a signal known to be due to sample mounting tape.  $R$ - Rayleigh surface mode, $LR$ - longitudinal surface resonance mode, $QT_i$ - bulk quasi-transverse acoustic phonon mode, $QL_i$ - bulk quasi-longitudinal acoustic phonon mode.}
    \label{fig:spectra}
\end{figure}

Fig. \ref{fig:spectra} shows representative spectra for each of the three samples, with additional spectra being presented in Fig. 1 of Ref. \cite{McNiven2022}.  In total, Brillouin peaks due to six distinct bulk acoustic phonon modes and two surface acoustic modes were observed.  These peaks have previously been assigned to four quasi-transverse sublattice modes ($QT_i$, $i=1-4$), two quasi-longitudinal sublattice modes ($QL_1$ and $QL_2$), the Rayleigh surface mode ($R$) and the longitudinal surface resonance ($LR$), the number and character of the bulk modes reflecting the incommensurate structure of Bi-2212 \cite{McNiven2022}.  The incommensurate structure is not manifested in the surface modes for which only the $R$ and $LR$ peaks were observed.  Moreover, the surface mode velocities $V_{R}$ are nearly independent of direction of propagation in the $ab$-crystallographic plane, while the bulk mode velocities showed little dependence on propagation direction over the range probed (as measured by angle from the crystallographic $c$-axis) \cite{McNiven2022}.  These velocities are presented in Table \ref{tab:velocities}.

\begin{table}[!htb]
\caption{Room-temperature surface and bulk phonon velocities (in m/s) for crystalline Bi-2212. Note: (i) ${V_{R}}_{(001)}$ is the average of Rayleigh mode velocities for crystals with $T_c = 90$ K and $T_c = 91$ K in the ab-plane; (ii) ${V_{L}}^{[001]}_{[001]}$ is the velocity of a bulk longitudinal mode that propagated along a direction $\sim5^\circ$ from the $c$-axis, while the remaining bulk mode velocities were measured for longitudinal modes that propagated at larger angles to this axis; (iii) bulk velocities are rounded to the nearest hundred due to an uncertainty of $\sim5\%$. See Ref. \cite{McNiven2022} for further details.}
\begin{ruledtabular}
 \begin{tabular}{c c c c c c c c c}
\multirow{2}{*}{${V_{R}}_{(001)}$}& \multirow{2}{*}{$V_{LR}$} & \multirow{2}{*}{$V_{QT_1}$}& \multirow{2}{*}{$V_{QT_2}$} & \multirow{2}{*}{$V_{QT_3}$} & \multirow{2}{*}{$V_{QT_4}$} & \multirow{2}{*}{$V_{QL_1}$} & \multirow{2}{*}{$V_{QL_2}$} & \multirow{2}{*}{${{{V_{L}}}^{[001]}_{[001]}}$}\\
&&&&&&&\\
\cline{1-9}
1590& 4260 & 2000 & 2400 & 3200 &4440 & 2700 & 6700 & 2700\\
\end{tabular}
\end{ruledtabular}
\label{tab:velocities}
\end{table}

\subsection{Elastic Constants}
\subsubsection{Influence of Incommensurate Structure \label{sec:IC}}
The elastic stiffness tensor for an incommensurate composite crystal with ``average'' orthorhombic symmetry is given by
\begin{equation}
    C_{ij}=
    \begin{pmatrix}
    C_{11} & C_{12} & C_{13} & 0  &0 & 0 \\
    C_{12} & C_{22} & C_{23} &  0 & 0& 0 \\
    C_{13}& C_{23} & C_{33} & 0  & 0& 0 \\
    0& 0 &0  & C_{44}  & 0& 0 \\
    0& 0 &0  & 0  & C_{55} & 0\\
    0& 0 & 0 & 0  &0 & C_{66} \\
    \end{pmatrix}.
\end{equation}
Diagonal elastic constants $C_{11}$, $C_{22}$, $C_{33}$, $C_{44}$, $C_{55}$, and $C_{66}$ can be determined directly from the density $\rho$ and phonon velocities in high symmetry directions via the following expressions \cite{Mah,Auld}:
\begin{equation}
\begin{split}
& C_{11}=\rho {\big({V_L}_{[100]}^{[100]}\big)}^2, \\
& C_{22}=\rho {\big({V_L}_{[010]}^{[010]}\big)}^2,\\
& C_{33}=\rho {\big({V_L}_{[001]}^{[001]}\big)}^2, \\
& C_{44}=\rho {\big({V_{TS}}_{[001]}^{[010]}\big)}^2 = \rho {\big({V_{TS}}_{[010]}^{[001]}\big)}^2, \\
& C_{55}=\rho {\big({V_{TF}}_{[001]}^{[100]}\big)}^2 = \rho {\big({V_{TS}}_{[100]}^{[001]}\big)}^2, \\
& C_{66}=\rho {\big({V_{TF}}_{[010]}^{[100]}\big)}^2 = \rho {\big({V_{TF}}_{[100]}^{[010]}\big)}^2,
\end{split}\label{eq:constants}
\end{equation}
where the subscript (superscript) denotes the propagation (polarization) direction and $V_{TS}$, $V_{TF}$ and $V_L$ are the slow transverse, fast transverse, and longitudinal acoustic phonon velocities, respectively.

In incommensurate composite crystals, at a critical wavevector $q_c$ (or equivalently, a critical frequency $f_{c_i}$ for each mode $i$), the bulk phonon dispersion exhibits a crossover from the familiar three-branch ($TS$, $TF$, and $L$) structure typical of a standard commensurate crystal to one containing multiple branches for each sublattice comprising the crystal.  For the particular case of two sublattices, the longitudinal bulk acoustic mode velocity in the incommensurate direction for $q\ll q_c$ ({\it i.e.}, the long-wavelength limit) is given by \cite{Finger1983} 
\begin{equation}
V_L^{q\rightarrow0}= \Bigg[\frac{m_A V_{L_A}^2+m_B V_{L_B}^2}{m_A + m_B} \Bigg]^{1/2},
\label{eqn:Vq0}
\end{equation}
where $m_\alpha$ and $V_{L_\alpha}$ ($\alpha=A, B$), are the sublattice masses and longitudinal phonon velocities, respectively.  This equation (Eq. \ref{eqn:Vq0}) has been used to determine $V_L^{q\rightarrow0}$ of BSCCO crystals in neutron scattering \cite{Etrillard2001,Etrillard2004} and Brillouin light scattering \cite{McNiven2022} experiments for which the crossover frequency from commensurate to incommensurate acoustic phonon dynamics lies in the range 10 MHz $< f_c <$ 10 GHz \cite{McNiven2022}. Ref \cite{Finger1983} implies that analogous relationships exist for transverse modes although these were not explicitly shown.

A particularly simple and useful relationship between crystal elastic constant $C_{22}$ and the associated sublattice constants can be derived by recalling the fundamental equation relating elastic modulus to acoustic phonon velocity and mass density, $C=\rho V^2$.  Substitution of this into Eq. \ref{eqn:Vq0} gives
\begin{equation}
 C_{22}=\rho_A V_{L_A}^2+\rho_B V_{L_B}^2 = C_{22}^A+C_{22}^B
\label{eqn:C_CA+CB}
\end{equation}
where $\rho_A$ and $\rho_B$ are the sublattice densities $\rho_\alpha=\rho m_\alpha / (m_A+m_B)$ and $V_{L_\alpha}$ ($\alpha=A, B$) are sublattice longitudinal acoustic phonon velocities in the incommensurate direction.  This result states that crystal elastic constant $C_{22}$ is simply the sum of the corresponding sublattice constants $C_{22}^A$ and $C_{22}^B$.  Thus, for an incommensurate composite system with ``average'' orthorhombic symmetry and incommensurate direction along the  crystallographic $b$-axis as for Bi-2212, knowledge of the velocities $V_{L_A}$ and $V_{L_B}$ along with sublattice densities allows $C_{22}$ to be determined.  

\subsubsection{$C_{11}$, $C_{33}$ and $C_{66}$} 
Elastic constants $C_{11}$, $C_{33}$, and $C_{66}$ were determined from the pertinent members of Eq. \ref{eq:constants}.  To obtain $C_{11}$ in this way, a value for ${V_L}_{[100]}^{[100]}$ is required.  For practical reasons, however, this velocity could not measured in the current experiments.  Instead, the velocity of the surface $LR$ mode (which is approximately equal to that of the bulk longitudinal mode \cite{caml1985}), $V_{LR} = 4260$ m/s, in an unspecified direction on the (001) plane was used.  The justification for this replacement is twofold: (i) the velocity of the $LR$ mode is essentially independent of direction on the (001) plane \cite{Boekholt} and thus the fact that the direction of propagation is not known is of no consequence, and (ii) the incommensurate structure of Bi-2212 is not manifested in the surface phonon dynamics and therefore surface phonons behave as they would in a typical commensurate crystal \cite{McNiven2022}.  This yields $C_{11} = 120$ GPa.  By a similar process, but more straightforward substitution, $C_{33} = 48$ GPa was obtained directly from the third member of Eq. \ref{eq:constants} using a velocity of 2720 m/s for a propagation direction very close ($< 5^\circ$) to the crystallographic $c$-axis for ${V_L}_{[001]}^{[001]}$.  The final member of Eq. \ref{eq:constants} gives $C_{66} = 40$ GPa with ${V_{TF}}_{[010]}^{[100]}=2460$ m/s from ultrasonics experiments on crystalline Bi-2212 at 270 K \cite{Wang2} as this velocity was not measured in the present work.  It is noted that for $270$ K $<T < 290$ K, the velocity of ultrasonic shear modes polarized and propagating in the $ab$-plane of quasi-crystalline Bi-2212 is nearly independent of temperature (a decrease of only $\sim10$ m/s from 270 K to 290 K), suggesting that the room temperature value of ${V_{TF}}_{[010]}^{[100]}$ in crystalline Bi-2212, and therefore $C_{66}$, is not appreciably different from its value at 270 K.  Moreover, extrapolation of the ${V_{TF}}_{[010]}^{[100]}$ versus temperature data for crystalline Bi-2212 \cite{Wang2} from $T>255$ K to room temperature, results in essentially the same velocity, and therefore $C_{66}$ value, as at 270 K due to its very weak temperature dependence.

\subsubsection{$C_{ii}^{\alpha}$ and $C_{22}$} \label{ss:C22}
Sublattice constants $C_{ii}^{\alpha} = \rho_{\alpha}V_{X_\alpha}^2$, where $\alpha = A, B$ and $X = L,\ TS,\ TF$ for $i = 2, 4, 6$, respectively, were determined from sublattice densities $\rho_{\alpha}$ and the velocities of the appropriate $QL_i$ and $QT_i$ modes from Table \ref{tab:velocities} for $V_{X_\alpha}$.  In particular, $V_{QL_1}$ and $V_{QL_2}$ are substituted for $V_{L_A}$ and $V_{L_B}$, the slow-quasi-transverse sublattice velocities $V_{QT_1}$ and $V_{QT_3}$ are substituted for $V_{TS_A}$ and $V_{TS_B}$ and, lastly, the fast-transverse sublattice velocities $V_{QT_2}$ and $V_{QT_4}$ are substituted for $V_{TF_A}$ and $V_{TF_B}$.  The rationale for replacing the the sublattice phonon velocities with $V_{QT_i}$ and $V_{QL_i}$ is provided in Ref. \cite{McNiven2022}.  The sublattice densities $\rho_A=4900\pm880$ kg$\cdot$m$^{-3}$ and $\rho_B=1700\pm300$ kg$\cdot$m$^{-3}$ were determined from $\rho_\alpha=\rho m_\alpha / (m_A+m_B)$ using a Bi-2212 density of $\rho=6600$ kg$\cdot$m$^{-3}$ and sublattice assignments of Bi$_2$Sr$_2$O$_4$ and CaCu$_2$O$_4$ for which $m_A = 658$ u and $m_B = 232$ u \cite{McNiven2022}. Numerical values for the sublattice constants are provided in Table \ref{tab:elastic}.

As per Eq. \ref{eqn:C_CA+CB}, elastic constant $C_{22} = 112\pm24$ GPa was obtained by adding sublattice constants $C_{22}^A$ and $C_{22}^B$.  While similar expressions to determine shear moduli from sublattice constants are not available in the literature as noted in Sec. \ref{sec:IC} and, furthermore, are anticipated to be somewhat more complicated than that for $C_{22}$ based on related calculations \cite{Axe1982, Finger1983}, values for $C_{44}^A$, $C_{44}^B$, $C_{66}^A$, and $C_{66}^B$ were computed based on the definition above and are provided in Table \ref{tab:elastic} in the event that they prove useful.

\subsubsection{$C_{12}$, $C_{44}$, $C_{23}$ and $C_{55}$}
Estimates of $C_{12}$, $C_{23}$, $C_{44}$, and $C_{55}$ were obtained from established relationships between select elastic constants and Rayleigh wave velocities $V_R$ in high symmetry directions on particular crystal planes \cite{Stoneley}. Specifically, these equations are:

\begin{equation}
\begin{split}
    \sqrt{1-\frac{\rho V_{R}^2}{C_{ii}}}\Bigg[1-\frac{C^2_{jk}}{C_{jj}C_{kk}}-\frac{\rho V_{R}^2}{C_{jj}}\Bigg] = \\ \sqrt{\frac{C_{jj}}{C_{kk}}} \Bigg[\frac{\rho V_{R}^2}{C_{jj}}\Bigg] \sqrt{1-\frac{\rho V_{R}^2}{C_{jj}}}.
    \end{split}
    \label{eq:stoneley} 
 \end{equation}
where the elastic constants subscripts for a particular direction of propagation and plane are given in Table \ref{tab:subscripts}. 

\begin{table}[h]
\caption{Elastic constants, identifying subscripts ($i, j, k$), and Rayleigh mode velocities ($V_{R}$) in Eq. \ref{eq:stoneley} for particular directions and planes of phonon propagation.}
\begin{ruledtabular}
\begin{tabular}{ccccc}
Direction & Plane & $V_{R}$ & $i, j, k$ & Elastic Constants \\ 
 & & [m/s] & & Appearing in Eq. \ref{eq:stoneley} \\ \hline
$\left [ 100\right ]$ & (010) & 1396\footnotemark[1] & 6, 1, 2 & $C_{66}$, $C_{11}$, $C_{22}$, $C_{12}$ \\
$\left [ 001\right ]$ & (010) & 1508\footnotemark[1] & 4, 3, 2 & $C_{44}$, $C_{33}$, $C_{22}$, $C_{23}$\\
$\left [010\right ]$ & (001) & 1590 & 4, 2, 3 & $C_{44}$, $C_{22}$, $C_{33}$, $C_{23}$ \\
$\left [100\right ]$ & (001) & 1590 & 5, 1, 3 & $C_{55}$, $C_{11}$, $C_{33}$, $C_{13}$
\end{tabular}
\end{ruledtabular}
\footnotetext[1]{From Ref. \cite{Boekholt}.}
\label{tab:subscripts}
\end{table}
 
Substituting the values of $C_{11}$, $C_{22}$, $C_{66}$, and ${V_{R}}_{[100]}^{(010)}$ into Eq. \ref{eq:stoneley} with $i, j, k = 6, 1, 2$ yields $C_{12}=102$ GPa.  Moreover, $C_{44}$ and $C_{23}$ are solved for simultaneously via Eqs. \ref{eq:stoneley} with $i, j, k = 4, 3, 2$ and $i, j, k = 4, 2, 3$ using $C_{22}$, $C_{33}$, ${V_{R}}_{[001]}^{(010)}$, and ${V_{R}}_{(001)}$ from Table \ref{tab:velocities} (noting the observed isotropy in the $ab$-plane \cite{McNiven2022,Boekholt}), which yield 20 GPa and 44 GPa, respectively.  Lastly, substituting $C_{11}$, $C_{33}$, and ${V_{R}}_{(001)}$, into Eq. \ref{eq:stoneley} with $i, j, k = 5, 1, 3$, and utilizing the approximation $C_{13} = C_{23}$, due to the fact that Bi-2212 is pseudo-tetragonal, gives $C_{55} = 20$ GPa. 

\begin{table*}[t]
\scriptsize
\caption{Room-temperature elastic constants (in GPa) and sublattice constants of crystalline Bi-2212 obtained in the present work and in previous studies.}
\begin{ruledtabular}
 \begin{tabular}{c c c c c c c c c c | c c c c c c }
\multirow{1}{*}{Technique/Study}& {$C_{11}$} & {$C_{22}$}& {$C_{33}$} & {$C_{44}$}& {$C_{55}$}& {$C_{66}$}& {$C_{12}$}& {$C_{13}$}& {$C_{23}$} & {$C^{A}_{22}$} & {$C^{B}_{22}$} & {$C^{A}_{44}$} & {$C^{B}_{44}$} & {$C^{A}_{66}$} & {$C^{B}_{66}$}\\[0.05 cm]
\hline
\multirow{1}{*}{Brillouin scattering - Pres} & 120$\pm$12 & 112$\pm24$ & 48$\pm5$ & 20$\pm$2 & 20$\pm$3 & -- & 101$\pm$16 & 44$\pm$8 & 44$\pm$8 & 36$\pm8$ & 76$\pm16$ & 19$\pm$4 & 17$\pm$4 & 25$\pm$5 & 33$\pm$7 \\[0.05 cm]

Brillouin scattering \cite{Boekholt} & 125$\pm$10  & -- & 76$\pm$9 & 16$\pm1$ & -- & -- & 79$\pm7$ & 56$\pm$6 & -- & -- & -- & -- & -- & -- & -- \\[0.05 cm]
Ultrasonics \cite{Wang1,Wang2}\footnotemark[1] & 134 & 114 & -- & -- & -- & 40 & -- & -- & -- & -- & -- & -- & -- & -- & -- \\[0.05 cm]
Neutron scattering \cite{Merritt2019}\footnotemark[2] & -- & -- & 39 & -- & -- & -- & -- & -- &-- & -- & -- & -- & -- & -- & --\\[0.05 cm]
Neutron scattering \cite{Etrillard2001}\footnotemark[3] & -- & -- & -- & -- & -- &  -- & -- & -- & -- & 15 & 139 & -- & -- & -- & --\\[0.05 cm]
Deformation Theory \cite{Jayachandran2002} & 118 & -- & 76  & 29 & --  & 37 & 40 & 26 & -- & -- & -- & -- & -- & -- & -- \\[0.05 cm]
Deformation Theory \cite{Jaya1999} & 119 & -- & 72  & 27 & --  & 40 & 40 & 27 & -- & -- & -- & -- & -- & -- & -- \\[0.05 cm]
\cline{1-16}
&&&&&&&&&&&\\
{\bf Best Estimate}  & {\bf 126$\pm$11}\footnotemark[4] & {\bf 113$\pm$24}\footnotemark[4] & {\bf 48$\pm$5} & {\bf 20$\pm$2} & {\bf 20$\pm$3}  & {\bf 40} & {\bf 101$\pm$17} & {\bf 44$\pm$8} & {\bf 44$\pm$8} & {\bf 36$\pm$8} & {\bf 76$\pm$16 } & {\bf 19$\pm$4} & {\bf 17$\pm$4} & {\bf 25$\pm$5} & {\bf 33$\pm$7} \\[0.3 cm]
\end{tabular}
\footnotetext[1]{Determined by estimating $V_{L,T}$ at 270 K from velocity vs temperature curves in the quoted study.}
\footnotetext[2]{Obtained by fitting $\omega=q{V_L}_{[001]}^{[001]}$ to experimental data provided in the quoted study.}
\footnotetext[3]{Estimated using velocities provided in the quoted study.}
\footnotetext[4]{Average of tabulated experimental values.}

\end{ruledtabular}
\label{tab:elastic}
\end{table*}

\section{Discussion}
Table \ref{tab:elastic} shows the elastic constants obtained in the present work and in previous ultrasonics \cite{Wang1,Wang2} and Brillouin scattering \cite{Boekholt} studies, as well as estimated values extracted from inelastic neutron scattering data \cite{Etrillard2001,Merritt2019} and calculated values from deformation theory \cite{Jayachandran2002,Jaya1999}.  Some of the elastic constants of the present work show excellent agreement with literature values while others are considerably different.  A critical comparison of values is given below with the aims of providing a set of best-estimate room-temperature elastic constants for Bi-2212, establishing the validity of the simple model outlined in Sec. \ref{sec:IC}, and evaluating the accuracy of the deformation theory calculations.

\subsection{{\boldmath $C_{11}$, $C_{22}$}: Confirmation of Published Values\label{ss:c11c22c66}}
Elastic constants $C_{11}$ and $C_{22}$ agree with previously published results within experimental uncertainty.  $C_{11} = 120\pm12$ GPa and $C_{22} = 112\pm24$ GPa lie within $\sim4$\% and $<2$\% of values obtained in previous Brillouin scattering \cite{Boekholt} and ultrasonics experiments \cite{Wang1,Wang2}, respectively.  It is also noted that the excellent agreement between $C_{11}$ of the present work and that reported in Ref. \cite{Boekholt}, both of which were determined from surface $LR$ mode velocities, is consistent with the finding that effects arising from the incommensurate structure are not manifested in the surface acoustic phonon dynamics of Bi-2212 \cite{McNiven2022}. 

\subsection{{\boldmath $C_{33}$, $C_{44}$, $C_{12}$, $C_{13}$}: Refinement of Published Values}
Elastic constants $C_{33}$, $C_{44}$, and $C_{12}$ of the current work differ considerably from published values.  In particular, the values of $C_{33}$, $C_{44}$, and $C_{12}$ are $\sim35$\% lower, $\sim20$\% higher, and $\sim30$\% higher than those reported in a previous Brillouin scattering study \cite{Boekholt}, respectively.  Comparison of the available data suggests that the values of $C_{33}$, $C_{44}$, and $C_{12}$ obtained in the current work most accurately reflect the true values of these constants.   

In the present study, $C_{33}$ was determined from the third member of Eq. \ref{eq:constants} using a velocity calculated from the frequency shift of a well-defined spectral peak arising from a bulk $QL$ mode propagating very nearly along [001].  The only other published value was that reported in a previous Brillouin study in which $C_{33}$ was estimated to be 76 GPa using a surface $LR$ mode velocity extracted from visibly lower quality spectra \cite{Boekholt}.  It is therefore likely that the value of $C_{33} = 48\pm5$ GPa obtained here more accurately reflects the true value of this constant.  Consistent with this conclusion is $C_{33}\approx40$ GPa from neutron scattering results on crystalline Bi-2212 in which the velocity was estimated by fitting $\omega=q{V_L}_{[001]}^{[001]}$ to a portion of the phonon dispersion curve constructed using data presented in Fig. 1a of Ref. \cite{Merritt2019}.  Moreover, ultrasonics experiments on highly-textured ceramic Bi-2212 with preferred grain orientation ({\it i.e.}, $c$-axis $\perp$ sample surface plane) yielded $C_{33}=44$ GPa \cite{Chang1993}.

The value of $C_{44} = 20 \pm 2$ GPa obtained from solving Eqs. \ref{eq:stoneley} simultaneously for $C_{44}$ and $C_{23}$ is in only fair agreement with the lone published experimental value of 16 GPa \cite{Boekholt}.  The latter, however, was determined using an approximation applicable to isotropic materials in which $V_{TS}$ is replaced with $V_R$ \cite{Boekholt}, and when used in Eq. \ref{eq:stoneley} along with $C_{22}$, $C_{33}$, and ${V_R}_{(001)}$, gives the unphysical result that $C_{23}$ is complex.


The high quality of the data used in determining $C_{12}$ via Eq. \ref{eq:stoneley} suggests that the value obtained in the present study is an accurate measure of this constant despite it being considerably higher than that obtained in previous Brillouin scattering experiments \cite{Boekholt}.  More specifically, the accuracy of the values of $C_{11}$, $C_{22}$ obtained here and used in Eq. \ref{eq:stoneley} has been verified by independent measurements as described in Sec. \ref{ss:c11c22c66}.  The value of  $C_{66}$ used in this equation, while not independently verifiable, was determined by straightforward calculation using the last member of Eq. \ref{eq:constants} with the velocity from ultrasonics measurements that, while measured at 270 K, does not differ significantly from its room temperature value due to it being nearly independent of temperature over the range 270 K to 290 K \cite{Wang2,Chang1993}.  In addition, the Rayleigh mode velocity required by Eq. \ref{eq:stoneley} was determined from a sharp, well-defined spectral peak resulting in a relatively low associated uncertainty \cite{Boekholt}.  Moreover, in Ref. \cite{Boekholt}, $C_{12}$ was calculated from questionable values of $C_{33}$ and $C_{44}$, as discussed in the preceding two paragraphs.

\subsection{{\boldmath $C_{55}$, $C_{23}$, $C_{66}$}: First Values} \label{first_values}
The values of $C_{55} = 20$ GPa, $C_{23} = 44$ GPa and $C_{66} = 40$ GPa are the first reported for these elastic constants at room temperature.  In the case of  $C_{55}$ and $C_{23}$, while there are no other published values to which to directly compare, one can invoke the fact that Bi-2212 is pseudo-tetragonal and compare to literature values of $C_{44}$ and $C_{13}$, respectively, to make a semi-quantitative statement on their accuracy.  

The value of $C_{55}$ obtained in the current study is quite close to that of $C_{44}$, as expected for a pseudo-tetragonal system.  Conversely, as mentioned above, $C_{23} = C_{13} = 44\pm8$ GPa. This value is $\sim20\%$ smaller than that reported previously for $C_{13}$ \cite{Boekholt}. As stated above, the value of $C_{13} = 56$ GPa from Ref. \cite{Boekholt} was determined using questionable values of $C_{33}=76$ GPa and $C_{44}=16$ GPa.

As described in detail in Sec. \ref{first_values}, the value of $C_{66}=40$ GPa should be robust as it was determined in a straightforward way from the last member of Eq. \ref{eq:constants} using an ultrasonics measurement on crystalline Bi-2212 at 270 K of a shear mode velocity that is, with high probability, virtually independent of temperature over the range 270 K to 290 K. Moreover, use of this value in calculations highlighted in the previous section yield values for other elastic constants that are in good agreement with those obtained in previous work.

\subsection{{\boldmath $C_{ii}^{\alpha}$}: Sublattice Elastic Constants}
Table \ref{tab:elastic} also reports sublattice elastic constants $C_{ii}^{\alpha}$, ($i=2,4,6$ and $\alpha = A, B$).  These are the only such values explicitly available in the literature and therefore direct comparison with the results of other studies was not possible.  Inspection of the $C_{22}^A$ and $C_{22}^B$ values obtained in the current work and those computed from available neutron scattering data \cite{Etrillard2001}, however, shows the former to be $\sim250$\% larger and $\sim50$\% smaller than the latter, respectively.  This large discrepancy suggests that at least one $C_{22}^A$-$C_{22}^B$ pair is incorrect.    Cognizant of this, the $C_{22}^A$ and $C_{22}^B$ values of the present work were added according to the simple model given by Eq. \ref{eqn:C_CA+CB} to yield $C_{22} = 112$ GPa.  This value is within $\sim4\%$ of a published value considered accurate \cite{Boekholt}.  In contrast, one obtains the seemingly unreasonably high value of $C_{22}=154$ GPa if the sublattice longitudinal mode velocities and densities based on the sublattice assignments in neutron scattering studies \cite{Etrillard2001} are used in Eq. \ref{eqn:C_CA+CB}.  Use of $V_{QL_1}$ and $V_{QL_2}$ from the present work with the sublattice densities from neutron scattering studies in Eq. \ref{eqn:C_CA+CB} gives an even higher value of $C_{22}=199$ GPa.

\subsection{Best Estimates}
The last row of Table \ref{tab:elastic} contains the best estimates of the room temperature elastic constants of Bi-2212 based on the above comparative analysis.  While most of the entries are those determined in the present work, the values of $C_{11}$ and $C_{22}$ obtained in the current study and in Refs.\cite{Boekholt,Wang1,Wang2} were considered equally accurate and therefore the average of these appears in the last row. 

The set of experimental best-estimate elastic constants in Table \ref{tab:elastic} can be compared to those obtained via deformation theory \cite{Jayachandran2002,Jaya1999}. Remarkably, the calculated values of $C_{11}$ and $C_{66}$ are within $<10\%$ of the corresponding measured values. Conversely, the calculated and experimental values of $C_{33}$, $C_{44}$, $C_{12}$, and $C_{13}$ are very different, possibly due to  exclusion of incommensurate structure effects in the calculations.

\subsection{Validity of Composite Incommensurate  Crystal Model}
The excellent agreement of elastic constant $C_{22}$, determined in this work by addition of the two associated sublattice constants, with a previously published value provides some measure of validation for the simple model outlined Sec. \ref{sec:IC}, the foundation of which is Eq. \ref{eqn:Vq0}.  It also suggests that the sublattice assignment in Ref. \cite{McNiven2022} is correct.  While an in-depth investigation is beyond the scope of this paper, intuitively, there appears to be no obvious reason why equations analogous to that used here to determine $C_{22}$ ({\it i.e.}, Eq. \ref{eqn:C_CA+CB}) could not be derived for other incommensurate composite crystals provided due consideration is given to crystal symmetry and the incommensurate structure.

\section{Conclusion}
Brillouin light scattering experiments were performed on crystalline  Bi-2212 at room-temperature yielding, through use of established and new relationships between phonon velocities and elastic moduli that take into account effects of incommensurate structure, a full set of best-estimate elastic constants and constituent sublattice constants. Elastic constant $C_{22}$, determined from the associated sublattice constants, showed remarkable agreement with a value obtained in previous work, providing support for the simple model linking this constant to sublattice constants and also for the sublattice assignment of Bi$_2$Sr$_2$O$_4$ and CaCu$_2$O$_4$ for Bi-2212 presented in Ref. \cite{McNiven2022}.  The data obtained in the present work could serve as a second check of the validity of this model if expressions relating other (particularly, shear) elastic constants or phonon velocities to sublattice constants or velocities were available in the literature.  More generally, the elastic constants presented in this work could aid in further development and refinement of models of elasticity and phonon dynamics in composite incommensurate systems.  In particular, the lack of agreement between several measured elastic constants provided in the present study and available theoretical values highlights the importance of extending elasticity theory to accurately connect the elastic properties of composite incommensurate crystals to those of the constituent sublattices.  A deeper understanding of the elastic properties and unusual acoustic phonon behaviour in Bi-2212 will also further understanding of electron-phonon coupling and bonding in layered cuprate superconductors and prove useful to those in the field of phonon engineering for the design and development of new acoustic and elasto/acousto-optic devices.

\section{Acknowledgements}
The authors would like to acknowledge Dr. J. P. Clancy at McMaster University, Canada, for supplying the samples used in this work. This work was partially funded by the Natural Sciences and Engineering Council of Canada (NSERC) through Discovery Grants to Andrews (\#RGPIN-2015-04306) and LeBlanc (\#RGPIN-2022-03882).
\bibliographystyle{apsrev4-1}
\bibliography{refs.bib}

\begin{thebibliography}{26}%
\makeatletter
\providecommand \@ifxundefined [1]{%
 \@ifx{#1\undefined}
}%
\providecommand \@ifnum [1]{%
 \ifnum #1\expandafter \@firstoftwo
 \else \expandafter \@secondoftwo
 \fi
}%
\providecommand \@ifx [1]{%
 \ifx #1\expandafter \@firstoftwo
 \else \expandafter \@secondoftwo
 \fi
}%
\providecommand \natexlab [1]{#1}%
\providecommand \enquote  [1]{``#1''}%
\providecommand \bibnamefont  [1]{#1}%
\providecommand \bibfnamefont [1]{#1}%
\providecommand \citenamefont [1]{#1}%
\providecommand \href@noop [0]{\@secondoftwo}%
\providecommand \href [0]{\begingroup \@sanitize@url \@href}%
\providecommand \@href[1]{\@@startlink{#1}\@@href}%
\providecommand \@@href[1]{\endgroup#1\@@endlink}%
\providecommand \@sanitize@url [0]{\catcode `\\12\catcode `\$12\catcode
  `\&12\catcode `\#12\catcode `\^12\catcode `\_12\catcode `\%12\relax}%
\providecommand \@@startlink[1]{}%
\providecommand \@@endlink[0]{}%
\providecommand \url  [0]{\begingroup\@sanitize@url \@url }%
\providecommand \@url [1]{\endgroup\@href {#1}{\urlprefix }}%
\providecommand \urlprefix  [0]{URL }%
\providecommand \Eprint [0]{\href }%
\providecommand \doibase [0]{http://dx.doi.org/}%
\providecommand \selectlanguage [0]{\@gobble}%
\providecommand \bibinfo  [0]{\@secondoftwo}%
\providecommand \bibfield  [0]{\@secondoftwo}%
\providecommand \translation [1]{[#1]}%
\providecommand \BibitemOpen [0]{}%
\providecommand \bibitemStop [0]{}%
\providecommand \bibitemNoStop [0]{.\EOS\space}%
\providecommand \EOS [0]{\spacefactor3000\relax}%
\providecommand \BibitemShut  [1]{\csname bibitem#1\endcsname}%
\let\auto@bib@innerbib\@empty
\bibitem [{\citenamefont {Wu}\ \emph {et~al.}(1993)\citenamefont {Wu},
  \citenamefont {Wang}, \citenamefont {Guo}, \citenamefont {Shen},
  \citenamefont {Yan},\ and\ \citenamefont {Zhao}}]{Wu1}%
  \BibitemOpen
  \bibfield  {author} {\bibinfo {author} {\bibfnamefont {J.}~\bibnamefont
  {Wu}}, \bibinfo {author} {\bibfnamefont {Y.}~\bibnamefont {Wang}}, \bibinfo
  {author} {\bibfnamefont {P.}~\bibnamefont {Guo}}, \bibinfo {author}
  {\bibfnamefont {H.}~\bibnamefont {Shen}}, \bibinfo {author} {\bibfnamefont
  {Y.}~\bibnamefont {Yan}}, \ and\ \bibinfo {author} {\bibfnamefont
  {Z.}~\bibnamefont {Zhao}},\ }\href {\doibase 10.1103/PhysRevB.47.2806}
  {\bibfield  {journal} {\bibinfo  {journal} {Phys. Rev. B}\ }\textbf {\bibinfo
  {volume} {47}},\ \bibinfo {pages} {2806} (\bibinfo {year}
  {1993})}\BibitemShut {NoStop}%
\bibitem [{\citenamefont {{J. Wu, Y. Wang, H. Shen and J. Zhu}}(1990)}]{Wu2}%
  \BibitemOpen
  \bibfield  {author} {\bibinfo {author} {\bibnamefont {{J. Wu, Y. Wang, H.
  Shen and J. Zhu}}},\ }\href {\doibase
  https://doi.org/10.1016/0375-9601(90)90591-B} {\bibfield  {journal} {\bibinfo
   {journal} {Phys. Lett. A}\ }\textbf {\bibinfo {volume} {148}},\ \bibinfo
  {pages} {127} (\bibinfo {year} {1990})}\BibitemShut {NoStop}%
\bibitem [{\citenamefont {{Y. Wang, J. Wu, J. Zhu, H. Shen, Y. Yan and Z.
  Zhao}}(1989)}]{Wang1}%
  \BibitemOpen
  \bibfield  {author} {\bibinfo {author} {\bibnamefont {{Y. Wang, J. Wu, J.
  Zhu, H. Shen, Y. Yan and Z. Zhao}}},\ }\href {\doibase
  https://doi.org/10.1016/0921-4534(89)91102-7} {\bibfield  {journal} {\bibinfo
   {journal} {Physica C}\ }\textbf {\bibinfo {volume} {162-164}},\ \bibinfo
  {pages} {454} (\bibinfo {year} {1989})}\BibitemShut {NoStop}%
\bibitem [{\citenamefont {{M. Saint-Paul and J.L.
  Tholence}}(1990)}]{Saint-Paul}%
  \BibitemOpen
  \bibfield  {author} {\bibinfo {author} {\bibnamefont {{M. Saint-Paul and J.L.
  Tholence}}},\ }\href {\doibase https://doi.org/10.1016/0921-4534(90)90035-D}
  {\bibfield  {journal} {\bibinfo  {journal} {Physica C}\ }\textbf {\bibinfo
  {volume} {166}},\ \bibinfo {pages} {405} (\bibinfo {year}
  {1990})}\BibitemShut {NoStop}%
\bibitem [{\citenamefont {{V. V. Aleksandrov, T. S. Velichkina, V. I.
  Voronkova, A. A. Gippius, S. V. Rek'ko, I. A. Yakovlev and V. K.
  Yanovskii}}(1990)}]{Aleksandrov}%
  \BibitemOpen
  \bibfield  {author} {\bibinfo {author} {\bibnamefont {{V. V. Aleksandrov, T.
  S. Velichkina, V. I. Voronkova, A. A. Gippius, S. V. Rek'ko, I. A. Yakovlev
  and V. K. Yanovskii}}},\ }\href {\doibase
  https://doi.org/10.1016/0038-1098(90)90116-S} {\bibfield  {journal} {\bibinfo
   {journal} {Solid State Commun.}\ }\textbf {\bibinfo {volume} {76}},\
  \bibinfo {pages} {685} (\bibinfo {year} {1990})}\BibitemShut {NoStop}%
\bibitem [{\citenamefont {{P. Baumgart, S. Blumenr{\"o}der, A. Erle, B.
  Billebrands, P. Splittgerber, G. G{\"u}ntherodt and H.
  Schmidt}}(1989)}]{Baumgart}%
  \BibitemOpen
  \bibfield  {author} {\bibinfo {author} {\bibnamefont {{P. Baumgart, S.
  Blumenr{\"o}der, A. Erle, B. Billebrands, P. Splittgerber, G. G{\"u}ntherodt
  and H. Schmidt}}},\ }\href@noop {} {\bibfield  {journal} {\bibinfo  {journal}
  {Physica C}\ }\textbf {\bibinfo {volume} {192-164}},\ \bibinfo {pages} {1073}
  (\bibinfo {year} {1989})}\BibitemShut {NoStop}%
\bibitem [{\citenamefont {{M. Boekholt, J.V. Harzer, B. Hillebrands and G.
  G{\"u}ntherodt}}(1991)}]{Boekholt}%
  \BibitemOpen
  \bibfield  {author} {\bibinfo {author} {\bibnamefont {{M. Boekholt, J.V.
  Harzer, B. Hillebrands and G. G{\"u}ntherodt}}},\ }\href {\doibase
  https://doi.org/10.1016/0921-4534(91)90017-S} {\bibfield  {journal} {\bibinfo
   {journal} {Physica C}\ }\textbf {\bibinfo {volume} {179}},\ \bibinfo {pages}
  {101} (\bibinfo {year} {1991})}\BibitemShut {NoStop}%
\bibitem [{\citenamefont {McNiven}\ \emph {et~al.}(2022)\citenamefont
  {McNiven}, \citenamefont {LeBlanc},\ and\ \citenamefont
  {Andrews}}]{McNiven2022}%
  \BibitemOpen
  \bibfield  {author} {\bibinfo {author} {\bibfnamefont {B.~D.~E.}\
  \bibnamefont {McNiven}}, \bibinfo {author} {\bibfnamefont {J.~P.~F.}\
  \bibnamefont {LeBlanc}}, \ and\ \bibinfo {author} {\bibfnamefont {G.~T.}\
  \bibnamefont {Andrews}},\ }\href {\doibase 10.1103/PhysRevB.106.054113}
  {\bibfield  {journal} {\bibinfo  {journal} {Phys. Rev. B}\ }\textbf {\bibinfo
  {volume} {106}},\ \bibinfo {pages} {054113} (\bibinfo {year}
  {2022})}\BibitemShut {NoStop}%
\bibitem [{\citenamefont {{J. Etrillard, Ph Bourges, H.F. He, B. Keimer, B.
  Liang and C.T. Lin}}(2001)}]{Etrillard2001}%
  \BibitemOpen
  \bibfield  {author} {\bibinfo {author} {\bibnamefont {{J. Etrillard, Ph
  Bourges, H.F. He, B. Keimer, B. Liang and C.T. Lin}}},\ }\href {\doibase
  10.1209/epl/i2001-00400-0} {\bibfield  {journal} {\bibinfo  {journal}
  {Europhys. Lett.}\ }\textbf {\bibinfo {volume} {55}},\ \bibinfo {pages} {201}
  (\bibinfo {year} {2001})}\BibitemShut {NoStop}%
\bibitem [{\citenamefont {He}\ \emph {et~al.}(2018)\citenamefont {He},
  \citenamefont {Wu}, \citenamefont {Song}, \citenamefont {Lee}, \citenamefont
  {Said}, \citenamefont {Alatas}, \citenamefont {Bosak}, \citenamefont
  {Girard}, \citenamefont {Souliou}, \citenamefont {Ruiz}, \citenamefont
  {Hepting}, \citenamefont {Bluschke}, \citenamefont {Schierle}, \citenamefont
  {Weschke}, \citenamefont {Lee}, \citenamefont {Jang}, \citenamefont {Huang},
  \citenamefont {Hashimoto}, \citenamefont {Lu}, \citenamefont {Song},
  \citenamefont {Yoshida}, \citenamefont {Eisaki}, \citenamefont {Shen},
  \citenamefont {Birgeneau}, \citenamefont {Yi},\ and\ \citenamefont
  {Frano}}]{He2018}%
  \BibitemOpen
  \bibfield  {author} {\bibinfo {author} {\bibfnamefont {Y.}~\bibnamefont
  {He}}, \bibinfo {author} {\bibfnamefont {S.}~\bibnamefont {Wu}}, \bibinfo
  {author} {\bibfnamefont {Y.}~\bibnamefont {Song}}, \bibinfo {author}
  {\bibfnamefont {W.-S.}\ \bibnamefont {Lee}}, \bibinfo {author} {\bibfnamefont
  {A.~H.}\ \bibnamefont {Said}}, \bibinfo {author} {\bibfnamefont
  {A.}~\bibnamefont {Alatas}}, \bibinfo {author} {\bibfnamefont
  {A.}~\bibnamefont {Bosak}}, \bibinfo {author} {\bibfnamefont
  {A.}~\bibnamefont {Girard}}, \bibinfo {author} {\bibfnamefont {S.~M.}\
  \bibnamefont {Souliou}}, \bibinfo {author} {\bibfnamefont {A.}~\bibnamefont
  {Ruiz}}, \bibinfo {author} {\bibfnamefont {M.}~\bibnamefont {Hepting}},
  \bibinfo {author} {\bibfnamefont {M.}~\bibnamefont {Bluschke}}, \bibinfo
  {author} {\bibfnamefont {E.}~\bibnamefont {Schierle}}, \bibinfo {author}
  {\bibfnamefont {E.}~\bibnamefont {Weschke}}, \bibinfo {author} {\bibfnamefont
  {J.-S.}\ \bibnamefont {Lee}}, \bibinfo {author} {\bibfnamefont
  {H.}~\bibnamefont {Jang}}, \bibinfo {author} {\bibfnamefont {H.}~\bibnamefont
  {Huang}}, \bibinfo {author} {\bibfnamefont {M.}~\bibnamefont {Hashimoto}},
  \bibinfo {author} {\bibfnamefont {D.-H.}\ \bibnamefont {Lu}}, \bibinfo
  {author} {\bibfnamefont {D.}~\bibnamefont {Song}}, \bibinfo {author}
  {\bibfnamefont {Y.}~\bibnamefont {Yoshida}}, \bibinfo {author} {\bibfnamefont
  {H.}~\bibnamefont {Eisaki}}, \bibinfo {author} {\bibfnamefont {Z.-X.}\
  \bibnamefont {Shen}}, \bibinfo {author} {\bibfnamefont {R.~J.}\ \bibnamefont
  {Birgeneau}}, \bibinfo {author} {\bibfnamefont {M.}~\bibnamefont {Yi}}, \
  and\ \bibinfo {author} {\bibfnamefont {A.}~\bibnamefont {Frano}},\ }\href
  {\doibase 10.1103/PhysRevB.98.035102} {\bibfield  {journal} {\bibinfo
  {journal} {Phys. Rev. B}\ }\textbf {\bibinfo {volume} {98}},\ \bibinfo
  {pages} {035102} (\bibinfo {year} {2018})}\BibitemShut {NoStop}%
\bibitem [{\citenamefont {{A. M. Merritt, J.-P. Castellan, T. Keller, S.R.
  Park, J.A. Fernandez-Baca, G.D. Gu and D. Reznik}}(2019)}]{Merritt2019}%
  \BibitemOpen
  \bibfield  {author} {\bibinfo {author} {\bibnamefont {{A. M. Merritt, J.-P.
  Castellan, T. Keller, S.R. Park, J.A. Fernandez-Baca, G.D. Gu and D.
  Reznik}}},\ }\href {\doibase 10.1103/PhysRevB.100.144502} {\bibfield
  {journal} {\bibinfo  {journal} {Phys. Rev. B}\ }\textbf {\bibinfo {volume}
  {100}},\ \bibinfo {pages} {144502} (\bibinfo {year} {2019})}\BibitemShut
  {NoStop}%
\bibitem [{\citenamefont {Andrews}(2018)}]{andr2018}%
  \BibitemOpen
  \bibfield  {author} {\bibinfo {author} {\bibfnamefont {G.~T.}\ \bibnamefont
  {Andrews}},\ }in\ \href@noop {} {\emph {\bibinfo {booktitle} {Handbook of
  Porous Silicon}}},\ \bibinfo {editor} {edited by\ \bibinfo {editor}
  {\bibfnamefont {L.}~\bibnamefont {Canham}}}\ (\bibinfo  {publisher} {Springer
  International},\ \bibinfo {year} {2018})\ Chap.~\bibinfo {chapter} {53}, pp.\
  \bibinfo {pages} {691--703}\BibitemShut {NoStop}%
\bibitem [{\citenamefont {Bozovic}(1990)}]{bozo1990}%
  \BibitemOpen
  \bibfield  {author} {\bibinfo {author} {\bibfnamefont {I.}~\bibnamefont
  {Bozovic}},\ }\href {\doibase 10.1103/PhysRevB.42.1969} {\bibfield  {journal}
  {\bibinfo  {journal} {Phys. Rev. B}\ }\textbf {\bibinfo {volume} {42}},\
  \bibinfo {pages} {1969} (\bibinfo {year} {1990})}\BibitemShut {NoStop}%
\bibitem [{\citenamefont {Hwang}\ \emph {et~al.}(2007)\citenamefont {Hwang},
  \citenamefont {Timusk},\ and\ \citenamefont {Gu}}]{hwan2007}%
  \BibitemOpen
  \bibfield  {author} {\bibinfo {author} {\bibfnamefont {J.}~\bibnamefont
  {Hwang}}, \bibinfo {author} {\bibfnamefont {T.}~\bibnamefont {Timusk}}, \
  and\ \bibinfo {author} {\bibfnamefont {G.}~\bibnamefont {Gu}},\ }\href
  {\doibase 10.1088/0953-8984/19/12/125208} {\bibfield  {journal} {\bibinfo
  {journal} {J. Phys. Condens. Matter}\ }\textbf {\bibinfo {volume} {19}},\
  \bibinfo {pages} {125208} (\bibinfo {year} {2007})}\BibitemShut {NoStop}%
\bibitem [{\citenamefont {Wang}\ \emph {et~al.}(2012)\citenamefont {Wang},
  \citenamefont {You}, \citenamefont {Xie}, \citenamefont {Lin},\ and\
  \citenamefont {Jiang}}]{wang2012}%
  \BibitemOpen
  \bibfield  {author} {\bibinfo {author} {\bibfnamefont {X.}~\bibnamefont
  {Wang}}, \bibinfo {author} {\bibfnamefont {L.~X.}\ \bibnamefont {You}},
  \bibinfo {author} {\bibfnamefont {X.~M.}\ \bibnamefont {Xie}}, \bibinfo
  {author} {\bibfnamefont {C.~T.}\ \bibnamefont {Lin}}, \ and\ \bibinfo
  {author} {\bibfnamefont {M.~H.}\ \bibnamefont {Jiang}},\ }\href {\doibase
  https://doi.org/10.1002/jrs.3120} {\bibfield  {journal} {\bibinfo  {journal}
  {J. Raman Spectrosc.}\ }\textbf {\bibinfo {volume} {43}},\ \bibinfo {pages}
  {949} (\bibinfo {year} {2012})}\BibitemShut {NoStop}%
\bibitem [{\citenamefont {Mah}\ and\ \citenamefont {Schmitt}(2003)}]{Mah}%
  \BibitemOpen
  \bibfield  {author} {\bibinfo {author} {\bibfnamefont {M.}~\bibnamefont
  {Mah}}\ and\ \bibinfo {author} {\bibfnamefont {D.~R.}\ \bibnamefont
  {Schmitt}},\ }\href {\doibase https://doi.org/10.1029/2001JB001586}
  {\bibfield  {journal} {\bibinfo  {journal} {J. Geophys. Res. Solid Earth}\
  }\textbf {\bibinfo {volume} {108}},\ \bibinfo {pages} {ECV 6} (\bibinfo
  {year} {2003})}\BibitemShut {NoStop}%
\bibitem [{\citenamefont {Auld}(1973)}]{Auld}%
  \BibitemOpen
  \bibfield  {author} {\bibinfo {author} {\bibfnamefont {B.~A.}\ \bibnamefont
  {Auld}},\ }\href@noop {} {\emph {\bibinfo {title} {Acoustic fields and waves
  in solids Vol 1}}}\ (\bibinfo  {publisher} {Wiley, New York},\ \bibinfo
  {year} {1973})\BibitemShut {NoStop}%
\bibitem [{\citenamefont {Finger}\ and\ \citenamefont
  {Rice}(1983)}]{Finger1983}%
  \BibitemOpen
  \bibfield  {author} {\bibinfo {author} {\bibfnamefont {W.}~\bibnamefont
  {Finger}}\ and\ \bibinfo {author} {\bibfnamefont {T.~M.}\ \bibnamefont
  {Rice}},\ }\href {\doibase 10.1103/PhysRevB.28.340} {\bibfield  {journal}
  {\bibinfo  {journal} {Phys. Rev. B}\ }\textbf {\bibinfo {volume} {28}},\
  \bibinfo {pages} {340} (\bibinfo {year} {1983})}\BibitemShut {NoStop}%
\bibitem [{\citenamefont {{J. Etrillard, L. Bourges, B. Liang, C.T. Lin and B.
  Keimer}}(2004)}]{Etrillard2004}%
  \BibitemOpen
  \bibfield  {author} {\bibinfo {author} {\bibnamefont {{J. Etrillard, L.
  Bourges, B. Liang, C.T. Lin and B. Keimer}}},\ }\href {\doibase
  10.1209/epl/i2003-10182-3} {\bibfield  {journal} {\bibinfo  {journal}
  {Europhys. Lett.}\ }\textbf {\bibinfo {volume} {66}},\ \bibinfo {pages} {246}
  (\bibinfo {year} {2004})}\BibitemShut {NoStop}%
\bibitem [{\citenamefont {Camley}\ and\ \citenamefont
  {Nizzoli}(1985)}]{caml1985}%
  \BibitemOpen
  \bibfield  {author} {\bibinfo {author} {\bibfnamefont {R.~E.}\ \bibnamefont
  {Camley}}\ and\ \bibinfo {author} {\bibfnamefont {F.}~\bibnamefont
  {Nizzoli}},\ }\href {\doibase 10.1088/0022-3719/18/24/023} {\bibfield
  {journal} {\bibinfo  {journal} {J. Phys. C: Solid State Phys.}\ }\textbf
  {\bibinfo {volume} {18}},\ \bibinfo {pages} {4795} (\bibinfo {year}
  {1985})}\BibitemShut {NoStop}%
\bibitem [{\citenamefont {Wang}\ \emph {et~al.}(1989)\citenamefont {Wang},
  \citenamefont {Wu}, \citenamefont {Zhu}, \citenamefont {Shen}, \citenamefont
  {Zhang}, \citenamefont {Yang},\ and\ \citenamefont {Zhao}}]{Wang2}%
  \BibitemOpen
  \bibfield  {author} {\bibinfo {author} {\bibfnamefont {Y.~N.}\ \bibnamefont
  {Wang}}, \bibinfo {author} {\bibfnamefont {J.}~\bibnamefont {Wu}}, \bibinfo
  {author} {\bibfnamefont {J.~S.}\ \bibnamefont {Zhu}}, \bibinfo {author}
  {\bibfnamefont {H.~M.}\ \bibnamefont {Shen}}, \bibinfo {author}
  {\bibfnamefont {J.~Z.}\ \bibnamefont {Zhang}}, \bibinfo {author}
  {\bibfnamefont {Y.~F.}\ \bibnamefont {Yang}}, \ and\ \bibinfo {author}
  {\bibfnamefont {Z.~X.}\ \bibnamefont {Zhao}},\ }\enquote {\bibinfo {title}
  {{U}ltrasonic study on anisotropic elasticity of
  {B}i$_2${S}r$_2${C}a{C}u$_2${O}$_8$ single crystal},}\ in\ \href@noop {}
  {\emph {\bibinfo {booktitle} {Beijing International Conference on High
  Temperature Superconductivity}}},\ Vol.~\bibinfo {volume} {22},\ \bibinfo
  {editor} {edited by\ \bibinfo {editor} {\bibfnamefont {Z.~X.}\ \bibnamefont
  {Zhao}}, \bibinfo {editor} {\bibfnamefont {G.~J.}\ \bibnamefont {Cui}}, \
  and\ \bibinfo {editor} {\bibfnamefont {R.~S.}\ \bibnamefont {Han}}}\
  (\bibinfo  {publisher} {World Scientific},\ \bibinfo {address} {Beijing},\
  \bibinfo {year} {1989})\ pp.\ \bibinfo {pages} {426--429}\BibitemShut
  {NoStop}%
\bibitem [{\citenamefont {Axe}\ and\ \citenamefont {Bak}(1982)}]{Axe1982}%
  \BibitemOpen
  \bibfield  {author} {\bibinfo {author} {\bibfnamefont {J.~D.}\ \bibnamefont
  {Axe}}\ and\ \bibinfo {author} {\bibfnamefont {P.}~\bibnamefont {Bak}},\
  }\href {\doibase 10.1103/PhysRevB.26.4963} {\bibfield  {journal} {\bibinfo
  {journal} {Phys. Rev. B}\ }\textbf {\bibinfo {volume} {26}},\ \bibinfo
  {pages} {4963} (\bibinfo {year} {1982})}\BibitemShut {NoStop}%
\bibitem [{\citenamefont {Stoneley}(1963)}]{Stoneley}%
  \BibitemOpen
  \bibfield  {author} {\bibinfo {author} {\bibfnamefont {R.}~\bibnamefont
  {Stoneley}},\ }\href {\doibase 10.1111/j.1365-246X.1963.tb06281.x} {\bibfield
   {journal} {\bibinfo  {journal} {Geophys. J. Int.}\ }\textbf {\bibinfo
  {volume} {8}},\ \bibinfo {pages} {176} (\bibinfo {year} {1963})}\BibitemShut
  {NoStop}%
\bibitem [{\citenamefont {Jayachandran}\ and\ \citenamefont
  {Menon}(2002)}]{Jayachandran2002}%
  \BibitemOpen
  \bibfield  {author} {\bibinfo {author} {\bibfnamefont {K.~P.}\ \bibnamefont
  {Jayachandran}}\ and\ \bibinfo {author} {\bibfnamefont {C.~S.}\ \bibnamefont
  {Menon}},\ }\href {\doibase 10.1088/0953-8984/14/1/306} {\bibfield  {journal}
  {\bibinfo  {journal} {J. Phys. Condens. Matter}\ }\textbf {\bibinfo {volume}
  {14}},\ \bibinfo {pages} {59} (\bibinfo {year} {2002})}\BibitemShut {NoStop}%
\bibitem [{\citenamefont {Jayachandran}\ and\ \citenamefont
  {Menon}(1999)}]{Jaya1999}%
  \BibitemOpen
  \bibfield  {author} {\bibinfo {author} {\bibfnamefont {K.~P.}\ \bibnamefont
  {Jayachandran}}\ and\ \bibinfo {author} {\bibfnamefont {C.~S.}\ \bibnamefont
  {Menon}},\ }\href {\doibase https://doi.org/10.1016/S0022-3697(98)00260-1}
  {\bibfield  {journal} {\bibinfo  {journal} {J. Phys. Chem. Solids}\ }\textbf
  {\bibinfo {volume} {60}},\ \bibinfo {pages} {267} (\bibinfo {year}
  {1999})}\BibitemShut {NoStop}%
\bibitem [{\citenamefont {Chang}\ \emph {et~al.}(1993)\citenamefont {Chang},
  \citenamefont {Ford}, \citenamefont {Saunders}, \citenamefont {Jiaqiang},
  \citenamefont {Almond}, \citenamefont {Chapman}, \citenamefont {Cankurtaran},
  \citenamefont {Poeppel},\ and\ \citenamefont {Goretta}}]{Chang1993}%
  \BibitemOpen
  \bibfield  {author} {\bibinfo {author} {\bibfnamefont {F.}~\bibnamefont
  {Chang}}, \bibinfo {author} {\bibfnamefont {P.~J.}\ \bibnamefont {Ford}},
  \bibinfo {author} {\bibfnamefont {G.~A.}\ \bibnamefont {Saunders}}, \bibinfo
  {author} {\bibfnamefont {L.}~\bibnamefont {Jiaqiang}}, \bibinfo {author}
  {\bibfnamefont {D.~P.}\ \bibnamefont {Almond}}, \bibinfo {author}
  {\bibfnamefont {B.}~\bibnamefont {Chapman}}, \bibinfo {author} {\bibfnamefont
  {M.}~\bibnamefont {Cankurtaran}}, \bibinfo {author} {\bibfnamefont {R.~B.}\
  \bibnamefont {Poeppel}}, \ and\ \bibinfo {author} {\bibfnamefont {K.~C.}\
  \bibnamefont {Goretta}},\ }\href {\doibase 10.1088/0953-2048/6/7/006}
  {\bibfield  {journal} {\bibinfo  {journal} {Supercond. Sci. Technol.}\
  }\textbf {\bibinfo {volume} {6}},\ \bibinfo {pages} {484} (\bibinfo {year}
  {1993})}\BibitemShut {NoStop}%
\end{thebibliography}%
\end{document}